\documentstyle [epsf]{l-aa}

\include{definizioni}
\begin{document}

\thesaurus{Section 06; 08.16.4, 08.05.3, 08.08.1} 

\title{Asymptotic Giant Branch predictions: theoretical uncertainties.}

\author{ S. Cassisi \inst{1}, V. Castellani \inst{2}$^,$ \inst{3}, S. Degl'Innocenti \inst{2}$^,$ \inst{3}, G. Piotto \inst{4}, M. Salaris \inst{5} }

\offprints {S. Degl'Innocenti, Dipartimento di Fisica Universit\`a 
di Pisa, piazza Torricelli 2, 56126 Pisa, Italy, scilla@astr18pi.difi.unipi.it}

\institute{
Osservatorio Astronomico di Collurania, via M. Maggini, I-64100 
Teramo, Italy
\and Dipartimento di Fisica, Universit\'a di Pisa, piazza Torricelli 2,
   I-56126 Pisa, Italy 
\and Istituto Nazionale di Fisica Nucleare, Sezione di Pisa, via
Livornese 582/A I-56010, S. Piero a Grado, Pisa, Italy
\and Dipartimento di Astronomia, Universit\'a di Padova, Vicolo
dell'Osservatorio 5, I-35122 Padova, Italy
\and Astrophysics Research Institute, Liverpool John Moores 
University, Twelve Quays House, Egerton Wharf, Birkenhead L41 1LD, UK
}

\date{Received 27 June 2000; accepted ..........   }

\maketitle

\markboth {Cassisi et al.: AGB predictions: theoretical uncertainties} {Cassisi et al.: AGB predictions: theoretical uncertainties}

\begin{abstract}

In this paper we investigate the level of agreement between
observations and ``new" Asymptotic Giant Branch (AGB) models, as
produced by updating the physical inputs adopted in previous stellar
computations. One finds that the new physics increases the predicted
luminosity of Horizontal Branch (HB) and AGB stellar structures by a
similar amount, keeping unchanged the predictions about the difference
in luminosity between these two evolutionary phases. The best fit of
selected globular clusters appears rather satisfactory, disclosing the
relevance of the assumption on the mass of the Red Giant Branch (RGB)
progenitor in assessing the distance modulus of moderately metal rich
clusters. The still existing uncertainties related either to the input
physics or to the efficiency of some macroscopic mechanisms, like
convection or microscopic diffusion, are critically discussed, ruling
out the occurrence of the so called ``breathing pulses" during the
central He exhaustion, in agreement with earlier suggestions.

\end{abstract}

\keywords{Stars:AGB, Stars: evolution, Stars: Hertzsprung-Russel diagram}

%__________________________________________________

\section{Introduction}

The capability of current stellar models to account for
all the evolutionary phases observed in stellar clusters is undoubtedly 
an exciting achievement  which crowns with success the development of 
stellar evolutionary theories as pursued all along the second half of the last
century. Following such a success, one is often tempted to use
evolutionary results in an uncritical way, i.e., taking these results at
their face values without allowing for theoretical uncertainties. 
However, theoretical uncertainties do exist, as it is clearly shown by the
not negligible differences still existing among evolutionary results 
provided by different theoretical groups. 

The discussion of these theoretical uncertainties was early addressed by
Chaboyer (1995) in a pioneering paper investigating the reliability of  
theoretical predictions concerning H-burning structures presently evolving 
in galactic globular clusters (GCs) and, in turn, on the accuracy of 
current predictions about GC ages. More recently, such an investigation 
has been extended to later phases of stellar evolution by Cassisi et al. 
(1998, hereinafter CCDW; 1999), and Castellani \& Degl'Innocenti (1999), 
who discussed theoretical predictions concerning central He-burning 
low-mass stars populating the Horizontal Branch of galactic
globular clusters.

In this paper we will discuss predictions concerning the evolutionary
behaviour of Asymptotic Giant Branch stars, devoting particular
attention to two key observational parameters, such as the luminosity
of the predicted AGB clump and the number ratio between HB and AGB
stars N$_{\mathrm AGB}$/N$_{\mathrm HB}$.  These parameters appear of particular relevance
since the AGB clump luminosity has been proposed as an alternative
distance indicator for old--intermediate age stellar populations
(Pulone 1992), while the ratio N$_{\mathrm AGB}$/N$_{\mathrm HB}$ is an excellent tool
for investigating the efficiency of mixing processes during the HB
phase (Buonanno et al. 1985), being HB lifetimes extremely sensitive
to the extension of the semiconvective region in the stellar core.\\
In the two next sections we will first discuss the differences between
``old" and ``new" models, testing the most updated theoretical scenario on
selected high-quality Color-Magnitude (CM) diagrams of galactic GCs.
Section 4 will deal with an investigation on the uncertainties still
existing in current theoretical models. Concluding remarks will close
the paper.

\section{Theoretical models.}

Theoretical predictions concerning AGB stars in galactic globulars
have been presented and discussed in a previous paper (``old'' models)
about a decade
ago (Castellani et al. 1991, hereinafter CCP). According to a quite
common procedure, the discussion was based on HB models as
produced by Red Giant Branch (RGB) progenitors with an original mass of 0.8M$_{\odot}$, 
in the assumption that, for ages of the order of 10$^{10}$ years, differences in ages 
play a minor role in defining the structure and the evolution of He-burning stars. 
On this basis it was shown that in clusters with a well populated red HB 
theory predicts the occurrence of a clump of AGB stars with a rather well 
defined low luminosity edge. Theoretical predictions were
found in reasonable agreement with the two main observational
parameters, as given by i) the luminosity of the AGB clump with respect to 
the HB and, ii) the number ratio of AGB to HB stars, the so-called $R_2$
parameter.
However, since that time evolutionary models have been 
progressively updated, following the availability of new and, hopefully, better
physics and - in particular - of better neutrino energy losses, equation of state,
opacities and  nuclear cross sections. 
Thus the problem arises if a good fitting is preserved 
even in recent models.

To address this question, Fig.~1 shows the time behaviour of
the luminosity for a typical HB model through and beyond the phase of central
He-burning, i.e., along both HB and AGB phases, as computed in CCP or with the updated 
theoretical scenario presented in CCDW (``new'' models), which takes also into account
the efficiency of element sedimentation in the RGB progenitors. 
The most evident difference is the decrease of the HB lifetime, 
already discussed in CCDW.
Consequently, the ratio of lifetimes in the 
two evolutionary phases, and thus the predicted star number ratio, is significantly
different, changing from $\tau$(AGB)/$\tau$(HB)$\sim$ 0.11 to about 0.15. 
As we will discuss later on in this paper, numerical experiments 
disclose that such differences in the He-burning lifetimes are largely due to both the 
decreased efficiency of the $^{12}$C($\alpha,\gamma$)$^{16}$O nuclear reaction 
in the ``new" models and to the change in radiative 
opacities. An additional, but secondary, contribution to the decrease of central He-burning 
lifetime follows the larger luminosity of the new ZAHB models.
However the same figure shows that the luminosity of both HB and AGB
clump is increased by quite a similar amount, so that the difference in luminosity 
between these two observables ($\Delta{M_V(AGB-HB)}$) as predicted in CCP, does 
survive the change of the physical inputs.

% FIGURA 1
\begin{figure}
\centerline{\epsfxsize= 8 cm \epsfbox{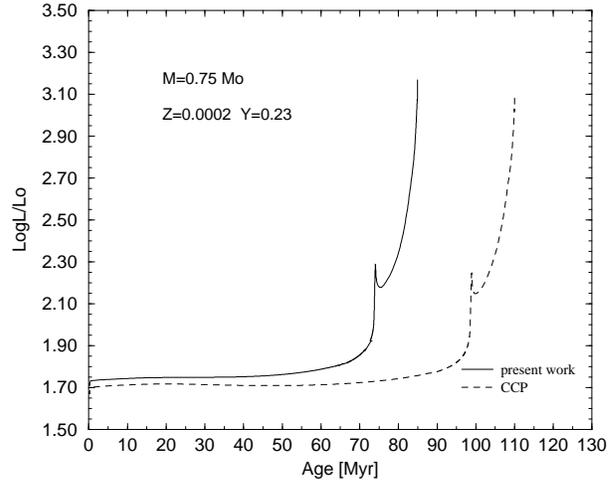}}
\label{tL}
\caption{Time behaviour of the surface luminosity during the central and
shell He-burning phases for a model with the labelled mass and
chemical composition, computed by adopting the most updated
physics (solid line) or as in CCP (dashed line).}
\end{figure}

\section {Observational tests.}

Taking advantage of the new and improved CM diagrams appeared in last
years, one can repeat the analysis already given in CCP to test the
adequacy of the theoretical scenario. To this purpose, we used
Castelli et al. (1997) model atmospheres to translate
bolometric luminosities and effective temperatures in the
observational $(M_V, B-V)$ CM diagram, constraining the mixing length
parameter by the requirement of reproducing the observed color of the
AGB branches.  In cool stars absolute visual magnitudes are indeed
dependent on the adopted efficiency of the external convection which
influences the effective temperature and, in turn, the bolometric
correction; thus meaningful predictions for the luminosity of the AGB
clump do require a suitable match of AGB colors.

Fig.~2 shows the best fit of present HB and post-HB evolutionary
tracks to the CCD CM diagram of M5 presented by Sandquist et al.
(1996). For the cluster metallicity we adopted from Sneden et
al. (1992) [Fe/H]$\approx$-1.17 with [$\alpha$/Fe]$\approx$+0.2.  By
adopting the relation given by Salaris et al. (1993),
and from the value of [$\alpha$/Fe] one derives Z=0.002 ([M/H]
$\approx$-1.03).  We also adopted Y=0.23. As expected, by keeping as
in CCP E(B-V)=0.03, in agreement also with the recent estimates by
Sandquist et al. (1996), one obtains a reasonable fit, provided that
the distance modulus estimated in CCP is increased by $\Delta(m-M)_V$
=0.09 mag. following the increased luminosity of the ``new'' models.
This distance modulus appears in good agreement with the value
provided by Sandquist et al. (1996, $(m-M)_V=14.50\pm0.07$ mag).
Sandquist et al. (1996) give for the observed number ratio
$R_2=N_{\mathrm AGB}/N_{\mathrm HB}$ the value $0.169\pm$0.06.  Within the uncertainty
this observational value appears consistent both with CCP
($\tau_{\mathrm AGB}/\tau_{\mathrm HB}\approx0.11$) and with present results
($\tau_{\mathrm AGB}/\tau_{\mathrm HB}\approx0.15$) even if the central value appears
in better agreement with the new models. However, in the next section
we will discuss the intrinsic weakness of such a theoretical
prediction.

%  FIGURA  2
\begin{figure}
\centerline{\epsfxsize=8 cm \epsfbox{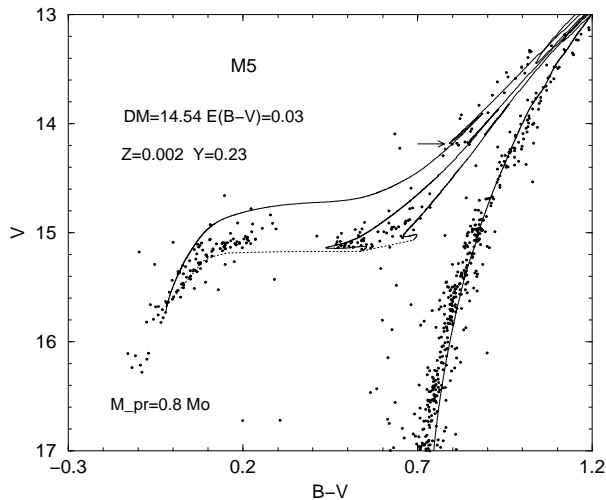}}
\label{M5}
 \caption{The observed CM diagram of M5 (Sandquist et al. 1996)
with superimposed selected evolutionary tracks of He burning models
with 0.8M$_\odot$ (Z=0.002 Y=0.23)  RGB progenitors.
The gap in the HB stellar distribution corresponds to the position of
the instability strip. The theoretical ZAHB is shown as a  dotted line.
The horizontal arrow marks the
predicted position of the low luminosity limit of the theoretical AGB clump.}
\end{figure}

\begin{table*}
\caption[]{Selected evolutionary quantities for Red Giant
models with different metallicity, initial mass and He content (see text for more details).
 The age at the He flash is in Gyr. 
}
\label{tab1}
\scriptsize
\begin{center}
\begin{tabular}{ c c c c c c c}
\hline
\hline\\
 $M$ (M$_\odot$)  &  Z & $Y_{\mathrm MS}$ & $Y_{\mathrm pred}$ & $\tau_{\mathrm flash}$(Gyr) & M$_{\mathrm cHe}$(M$_\odot)$ & 
Y$_{\mathrm HB}$ \\
\hline
0.80   &  0.0002 &  0.230    &  0.230    &  12.8   & 0.5148  &  0.2261  \\
0.80   &  0.002  &  0.230    &  0.234    &  15.4   & 0.5041  &  0.2305  \\
0.80   &  0.006  &  0.230    &  0.242    &  19.9   & 0.5001  &  0.2308  \\
0.80   &  0.006  &  0.242    &  0.242    &  18.3   & 0.4967  &  0.2457   \\
0.90   &  0.006  &  0.270    &  0.242    &  10.8   & 0.4892  &  0.2782   \\
0.95   &  0.006  &  0.230    &  0.242    &  11.0   & 0.4936  &  0.2420   \\
0.95   &  0.006  &  0.242    &  0.242    &  10.0   & 0.4933  &  0.2558   \\
\hline 
\end{tabular}
\end{center}
\end{table*}

To test theoretical models at larger metallicities, the same
fitting procedure has been applied to the high-quality CM diagram for
47 Tuc (Sosin et al. 1996), as obtained with the WFPC2 camera of the
Hubble Space Telescope.  For the cluster metallicity, we adopted the
spectroscopical measurement by Carretta \& Gratton (1997)
$[Fe/H]=-0.7$
and an $\alpha$-element enhancement
$[\alpha/Fe]\approx 0.2$ which corresponds to a mean between the
estimates listed by Carney (1996) and by Salaris \& Cassisi
(1996). Thus again from Salaris et al. (1993)
one obtains Z=0.006. According to Schlegel et al.  (1998)  a
 cluster reddening E(B-V)=0.03 has been adopted.

% FIGURA  3 

\begin{figure}
%\hspace{2cm}
\centerline{\epsfxsize=  8 cm \epsfbox{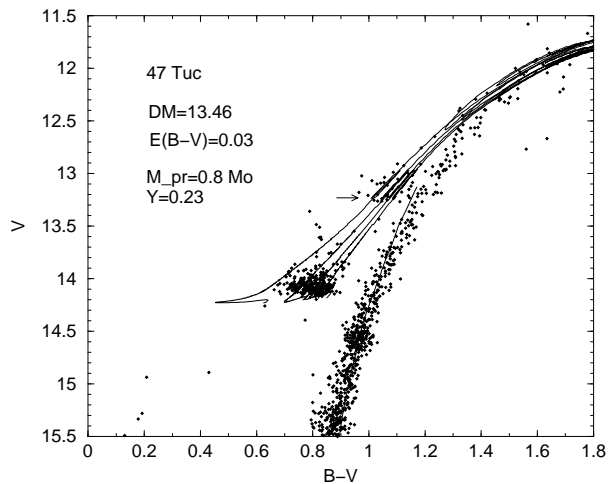} }
\caption{ The observed CM diagram of 47 Tuc (Sosin et al. 1996) with
superimposed selected evolutionary tracks of He-burning models with a
0.80 M$_{\odot}$ (Z=0.006 Y=0.23) RGB progenitor.
The ZAHB models are fitted to the observed lower envelope of the HB distribution for
the labelled assumptions on the distance modulus and reddening. The
horizontal arrow marks the predicted lower envelope of the AGB clump.}
\end{figure}

With this choice Fig. 3 shows that the fitting of HB and AGB stars for
47 Tuc keeps being rather satisfactory.  The derived distance modulus
appears in agreement with the one obtained by Salaris \& Weiss (1998,
$(m-M)_V=13.42 \div 13.50$ mag.) on theoretical basis, but smaller
than the empirical value obtained by Gratton et al. (1997,
$(m-M)_V=13.62\pm0.08$ mag) on the basis of Hipparcos subdwarfs.

% FIGURA 4
%
\begin{figure}
\hspace{2cm}
\centerline{\epsfxsize=  8 cm \epsfbox{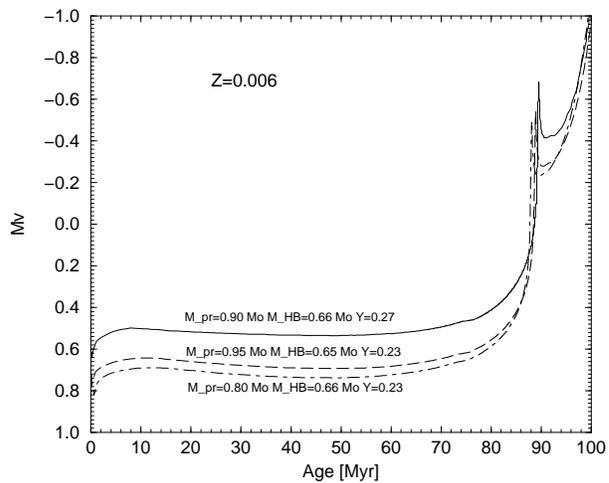} }
\caption{Time behaviour of the absolute visual magnitude for He-burning models
with selected RGB progenitors from the ZAHB to the AGB (see text).}
\end{figure}

As already mentioned, in the previous fit of 47Tuc we used the common
procedure to adopt a 0.8 M$_\odot$ (Y=0.23) RGB progenitor.  However,
for moderately metal rich clusters this assumption is not completely
satisfactory. To discuss this point Table 1 shows selected
evolutionary parameters for Red Giant models with different
metallicity, initial mass and He content. Left to right one finds: the
mass of the RGB progenitor, the adopted values of Z and initial Helium
abundance (Y$_{\mathrm MS}$), the value of Y predicted (Y$_{\mathrm
pred.}$) when accounting for a galactic Helium enrichment
$\Delta{Y}/\Delta{Z}\approx2.4$ (see, e.g., Pagel \& Portinari 1998,
Castellani et al. 1999), the age at the He flash ($\tau_{\mathrm
flash}$) together with the size (M$_{\mathrm cHe}$) of the He core and
the amount of surface He ($Y_{\mathrm HB}$) at this stage. One notices
that the assumptions about the original He ($Y_{\mathrm MS}=0.23$) and
the mass of the RG progenitor (M=0.8 M$_{\odot}$) provide cluster ages
which appear reasonably adequate for metal-poor globulars, with
Z$\approx$0.0002.  However, when increasing the metallicity up to
Z=0.006, the same assumptions would imply an exceedingly large cluster
age and an amount of original He not negligibly lower than expected by
assuming a reasonable value for the galactic correlation between Y and
Z. The same table shows that an RGB progenitor of 0.95 M$_{\odot}$
would give an age at the He flash in better agreement with present
estimates for this cluster (Gratton et al. 1997, Salaris \& Weiss
1998).

Fig.~4 shows the temporal behaviour of the visual magnitude for He
burning models with Z=0.006 and selected RGB progenitors.  The mass of
the HB models has been chosen to approximately fit the center of the
observational horizontal branch of 47 Tuc. As a relevant point, Fig.4
shows that increasing the mass of the RGB progenitor (but adopting the
same initial He abundance), the HB and AGB luminosity levels increase
again by a rather similar amount. Thus one can easily predict that
decreasing the age theory will fit the two He burning phases with the
same accuracy, but with a distance modulus increased by $\Delta
(m-M)_V$ $\approx$0.06. From the same figure one can estimate that
passing from Y=0.23 to Y=0.242 one expects a further increase by about
0.04 mag. As a result, we drive the attention on the evidence that the
commonly used assumptions for an RGB progenitor with M=0.8 M$_{\odot}$
and Y=0.23 can underestimate the cluster distance modulus by about $\Delta
(m-M)_V$ $\approx$0.1.  With such a correction, now one finds that
our theoretical predictions appear in excellent agreement with the
previous quoted estimate by Gratton et al. (1997).

However, regarding the helium content, Salaris \& Weiss (1998) noted
that there are in literature some suggestions for an helium abundance
of 47Tuc close to the solar value.  The same Fig.4 shows a 0.66 M$_{\odot}$
HB model with a 0.9 M$_{\odot}$ Y=0.27 RGB progenitor. The
age at the He flash is about 11 Gyr, in agreement with recent age
estimates for 47 Tuc.  One finds that now the HB luminosity is
increased by about 0.23 mag. with respect to the model with
M=0.8 M$_{\odot}$ Y=0.23 progenitor whereas, even with this huge He
variation, the $\Delta{M_V(AGB-HB)}$ parameter is preserved with an
accuracy of few hundredth of magnitude.  Thus the observed
$\Delta{M_V(AGB-HB)}$ cannot constrain the amount of original He.  Nor
the suggestion for an initial He abundance Y$\approx$0.27 
is ruled out by the luminosity of the RGB bump,
as disclosed by numerical experiments.

Before closing this section one has to discuss briefly the
reasons why $\Delta{M_V(AGB-HB)}$ appears largely independent of
variations in the assumed progenitor mass and/or original helium
content. The latter point was already discussed by Pulone (1992),
starting from the well known evidence that an increase in the original
He content produces ZAHB models with smaller initial He-cores, but
larger luminosities. The initial AGB luminosity is however also
larger, since during HB evolution stars with higher helium abundance
burn their hydrogen with larger rates, so that increasing the original
He also increases the He-core at the He-exhaustion.\\ 
As for the behaviour of $\Delta{M_V(AGB-HB)}$ with the progenitor mass, an
increase of the RGB progenitor mass produces a slight decrease of the
He core mass at the He ignition, and an increase of the envelope He
abundance. In canonical models this is a consequence of a larger efficiency
of the I$^o$ dredge up. When element diffusion is taken into account this
also arises from the evidence that larger masses have shorter evolutionary times
and the efficiency of diffusion is thus reduced (see e.g. Proffit \&
VandenBerg, 1991, Castellani \& Degl'Innocenti, 1999).
According to the previous discussion, the competitive 
effects of these occurrences on the ZAHB and AGB clump luminosity, make
almost constant the value of $\Delta{M_V(AGB-HB)}$.

\section{Theoretical uncertainties}

In this section we will refer to the ``new" evolutionary scenario to
discuss the uncertainties affecting this as any other current
theoretical prediction.  We will separately discuss uncertainties
produced by macroscopic mechanisms or by intrinsic uncertainties in
the adopted input physics.

\begin{table*}
\begin{center}
\caption{Selected results for a 0.75 M$_{\odot}$ He-burning model with a 0.8 M$_{\odot}$ RGB 
progenitor and metallicity  Z=0.0002, under different assumptions about the adopted physical inputs.
All the evolutionary lifetimes are in $10^6$ yrs.}
\scriptsize
\begin{tabular}{c c c c c c c c c c c}
\hline
\hline\\
    & $\tau_{0.1}$  & $\tau_{\mathrm HB}$   &  $\tau_{\mathrm AGB}$  &  $\tau_{\mathrm AGB}/\tau_{\mathrm HB}$ & 
    $\log(L/L_{\odot})^{\mathrm AGB}$  & M$_{\mathrm CO}^{He}$(M$_\odot)$ & M$_{\mathrm CO}^{\mathrm AGB}$(M$_\odot)$  &  X$_{\mathrm C}$  &   
X$_{\mathrm O}$  &  X$_{\mathrm C}$/X$_{\mathrm O}$  \\
\\
\hline
reference               & 65.3 & 73.6   &   11.30  &  0.154   &   2.178 &  
0.215  & 0.505 & 0.514  &  0.486 & 1.058 \\
old cross sections   & 72.4 & 81.4   &   10.95  &  0.134   &   2.173 &  
0.217  & 0.505 & 0.240  &  0.760 & 0.316 \\
EOS Straniero 1988   & 73.6 & 83.5   &   11.02  &  0.132   &   2.146 &  
0.206  & 0.499 & 0.237  &  0.763 & 0.311 \\
LAOL opacity         & 80.9 & 90.9   &   10.97  &  0.121   &   2.148 &  
0.221  & 0.506 & 0.230  &  0.769 & 0.299 \\
old plasma neutrinos & 80.9 & 90.8   &   11.00  &  0.121   &   2.150 &  
0.216  & 0.505 & 0.230  &  0.770 & 0.299 \\
CCP                  & 87.4 & 98.4   &   11.00  &  0.112   &   2.148 &  
0.213  & 0.495 & 0.225  &  0.775 & 0.290 \\  
\hline
\end{tabular}
\end{center}
\end{table*}

%  FIGURA 5

\begin{figure}
\centerline{\epsfxsize=  8 cm \epsfbox{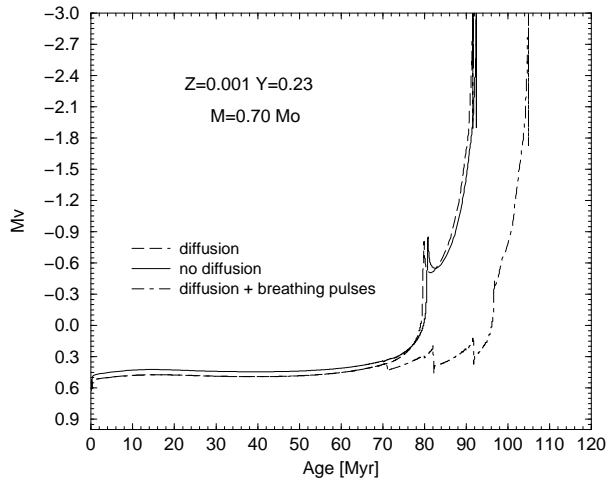}}
\caption{Time behaviour of the absolute visual magnitude for a 0.7M$_{\odot}$ He-burning model 
from the ZAHB until the first thermal pulse by accounting
(dashed line) or neglecting (solid line) atomic diffusion in the computation of the 0.8M$_{\odot}$
RGB progenitor. The dot-dashed line corresponds to the 0.7M$_{\odot}$ He-burning
model with diffusion when breathing pulses are allowed to occur during the central
He-burning evolutionary phase.}
\end{figure}

\subsection{Macroscopic mechanisms}
 
Any evaluation of HB and AGB models has to be based, implicitly or
explicitly, on suitable assumptions about the efficiency of some
macroscopic mechanisms. When dealing with low mass stars, one has to properly take into account:
i) the amount of mass loss, ii) the
efficiency of element sedimentation and, last but not least, iii) the
amount of convective mixing in the stellar interior and, in this
context, the debated occurrence of the so called "breathing pulses"
(Sweigart \& Demarque 1972, 1973; Castellani et al. 1985, Sweigart 1990 and references
therein). 
The amount of mass loss is constrained by the observed colour HB
distribution, and it can be reasonably taken into account when fitting
clusters CMD by reproducing the HB colour. 
On the contrary, the other two mechanisms have much
more subtle effects, worth to be investigated in some details.

As for element diffusion, Fig.~5 compares the behaviour with time of
the absolute visual magnitude for a typical HB model (M=0.7
M$_{\odot}$) with or without microscopic diffusion in the H-burning
progenitor. As already discussed (see, e.g. CCDW, Castellani \&
Degl'Innocenti 1999) the difference is small: if diffusion is not
taken into account the luminosity of both the HB and AGB models slightly increases
by the same amount. Thus for no diffusion models the estimate of the distance modulus
increases by about $\Delta{M_V}\approx$0.05 mag. In passing, we note that the
evolutionary time of both the HB and AGB phases remains practically
unchanged.

As shown in the same Fig.~5, this is not the case for the model with
breathing pulses (which are suppressed in our reference ``new''
models), since the HB lifetime is increased by more than 20\% whereas
the AGB lifetime is decreased by more than 25\%. As a result the ratio
$\tau_{\mathrm AGB}/\tau_{\mathrm HB}$ is dramatically reduced with respect to
standard calculations. For models suitable for AGB stars in M5 the
inclusion of breathing pulses would produce a ratio $R_2=
N_{\mathrm AGB}/N_{\mathrm HB}\sim0.08$, well below the range of values allowed by
observational constraints.  In addition, Fig.~5 shows that if
breathing pulses are at work, one expects a fainter and proportionally
less populated AGB phase, with a less evident clumping of AGB stars at
the bottom of the AGB branch.  The occurrence of breathing pulses has
been largely debated in the literature. Caputo et al. (1989) compared
HB and RGB evolutionary lifetimes with observational data to conclude
for the inefficiency of the pulses.  One easily finds that present
results clearly run against observations, thus reinforcing the above
quoted suggestion for the inefficiency of this phenomenon.

%FIGURE 6

\begin{figure}
\centerline{\epsfxsize=  8 cm \epsfbox{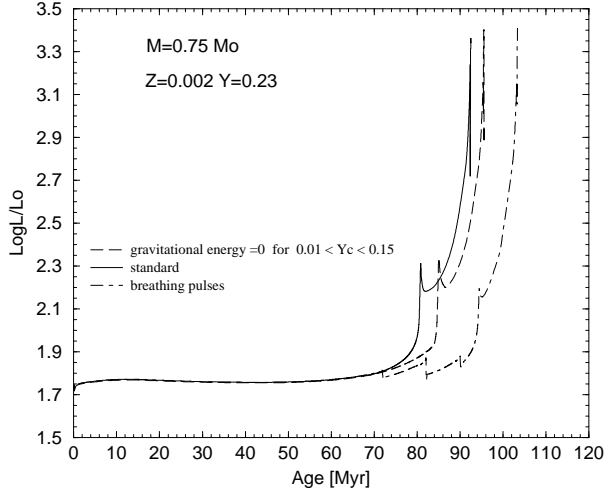}}
\caption{Time behaviour of the surface luminosity for a 0.75 M$_{\odot}$ He-burning model from
the ZAHB until the first thermal pulse when breathing pulses are
included (dot-dashed line).  The figure also shows the same model
when breathing pulses are dampened by adopting two different numerical
algorithms: by preventing any He increase in the central convective
core (solid line) or by neglecting the gravitational
energy release during the phase of He exhaustion (dashed line, see text
for more details).
In all numerical experiments, microscopic diffusion 
of both He  and heavy elements has been accounted for in the 0.8M$_{\odot}$ RGB 
progenitor.}
\end{figure}

However, different algorithms to avoid the occurrence of breathing
pulses in computing models have been presented in the literature. As
discussed in Chieffi \& Straniero (1989), our evolutionary code just
prevents any increase of He abundance in the central convective cores
during the HB evolutionary phase, whereas Dorman \& Rood (1993) have
shown that breathing pulses can be suppressed by neglecting the
generation of gravitational energy during the core-helium exhaustion
phase. Fig.~6 shows the time behaviour of the luminosity of a
0.75 M$_{\odot}$ He-burning model for these two alternative assumptions
about breathing pulses suppression.  One finds that neglecting the
gravitational energy produces longer central He-burning lifetimes and
slightly brighter AGB than our approach does. 
This because these models extend the semiconvection in a
larger region than our standard models do.
Even if the differences appear rather marginal, difficult to be
detected within the current observational uncertainties, one should
not forget that the way of suppressing the breathing pulses alone
affects the theoretical models with an uncertainty in the HB lifetime
of the order of 5\%, and of about 0.08 mag. in the absolute magnitude
of the lower envelope of the AGB clump. In this context, we are
actually reluctant to make a choice between the two quoted
mechanisms. As a matter of fact, if our approach is rather crude,
it is also true that gravitational energy is at the work in real
stars. Thus the above quoted uncertainties appear to us unavoidables.

\subsection{Uncertainties in input physics.}

In the previous Fig.~1 we have already shown the not negligible
differences between present and CCP results, as produced by updating
the physical inputs. To discuss these differences in more details we
report in Table~2 selected quantities depicting evolutionary
predictions for a 0.75M$_{\odot}$ He-burning model (Z=0.0002, Y= 0.23)
under various assumptions about the input physics. Top to bottom the
first line gives selected results from the present ``new'' reference
scenario (with element diffusion) and, below, predictions for the same
model but with progressive variations in the input physics eventually
reproducing the input of CCP models. In all cases, but the
``CCP'' one, the helium core mass and the envelope helium abundance in
the HB phase are those of the ``new'' reference model, that is the
changes in the physical inputs are applied only to the HB evolution.
The variations in the physics of the reference model run, top to
bottom, as follows: i) nuclear cross sections from Caughlan \& Fowler
(1988) to Caughlan et al. (1985) ii) as in previous model but with the
Equation of State of Straniero (1988) instead of the OPAL EOS (Rogers
1994, Rogers et al. 1996), iii) as in the previous model but with
radiative opacity from Livermore opacity tables (Iglesias \& Rogers
1996) to Los Alamos opacities (Huebner et al. 1977), and electron
conductivity from Itoh et al. (1983) to Hubbard \& Lampe (1969), iv)
as in the previous model but passing from plasma neutrino energy
losses of Haft, Raffelt \& Weiss (1994) to those of Munakata et
al. (1985).  The last row gives the original CCP result as computed
taking also into account the effects of the ``old'' physics on the RGB
progenitor, that is by adopting as helium core mass and surface
helium abundance the values by CCP.

Left to right the columns give: the time spent in the central He
burning to reach a central He abundance $Y_c=0.1$ (corresponding to
the onset of breathing pulses, which are anyway suppressed), the total
He-burning lifetime $\tau_{\mathrm HB}$, the AGB lifetime ($\tau_{\mathrm AGB}$, from
the exhaustion of central He to the onset of thermal pulses), the
ratio $\tau_{\mathrm AGB}/\tau_{\mathrm HB}$, the bottom luminosity of the AGB clump,
the mass of the CO core at the He exhaustion and at the first thermal
pulse and, finally, the fractional abundance by mass of C and O in the
CO core at the central He exaustion and the ratio of these two
quantities.
 
One easily notices that the He-burning lifetime is mainly influenced
by  nuclear cross sections and  by radiative opacity,  
which both affect the lifetime by about 9-10\%. As for
nuclear cross sections, the main variation is due to the increase 
by about a factor 2 of the $^{12}$C + $\alpha$ cross section 
from Caughlan \& Fowler 1988 to Caughlan  et al. (1985).
The opacity effect is almost completely due to variations in 
carbon-oxygen opacity, which affects the opacity in the stellar core. 
One should also note the dramatic effect of the variation of $^{12}$C +$\alpha$ nuclear cross 
section on the C/O ratio in the core, a quantity which is governing the subsequent white 
dwarf evolution (see e.g.  D'antona \& Mazzitelli 1990, Wood 1992,
Salaris et al.~1997). Bearing in mind
that a realistic estimate of current uncertainties affecting this
cross section is as large as a factor of two (see Caughlan \& Fowler
1988, Buchmann 1996, Angulo et al. 1999), more precise measurements of
this cross section at energies of astrophysical interest appear of
great relevance.

\section{Conclusions}

In this paper we found that the ``new" theoretical scenario arising
from stellar models with updated physical inputs, appears able to
account for the main observational constraints on GC AGB
stars. However, we insist on the evidence that no theoretical result
can be taken at its face value, because of still existing theoretical
uncertainties. In this context we critically discussed the
indetermination due either to the efficiency of some macroscopic
mechanisms, like atomic diffusion and breathing pulses, or to the
uncertainties on the input physics (equation of state, opacity,
nuclear cross sections etc.).  While the efficiency of atomic
diffusion has little influence on He-burning models, the occurrence of
breathing pulses is ruled out by the comparison between theory and
observation, confirming the conclusion reached by Caputo et al. (1989)
in the frame of the ``old" theoretical scenario.

Regarding the physical inputs, the main uncertainty still present in
He-burning models is given by the $^{12}C(\alpha,\gamma)^{16}O$ cross
section, which influences the He burning lifetimes and the C/O ratio
in the carbon-oxygen core, with relevant consequences on the final
cooling of white dwarfs.  Since this nuclear reaction rate has a
strong effect on the predicted central He burning lifetimes, it
affects also the evaluation of the initial He abundance in galactic
GCs via the $R$ parameter, i.e., the ratio between the HB stars and
the RGB ones brighter than the HB (see e.g. the discussion in CCDW and
in Zoccali et al. 2000). On the contrary, data in Table~2 show that
AGB lifetimes appear marginally affected by this uncertainty;
therefore one could be tempted to use the ratio N$_{\mathrm AGB}$/N$_{\mathrm HB}$ as
an indicator of efficiency of the $^{12}$C +$\alpha$ reaction.
However, as discussed in the previous section, the extension of the
mixed core of He-burning stars affects the value of N$_{\mathrm AGB}$/N$_{\mathrm HB}$
too, so that a comparison with a given CMD can only tell us if the
combination of mixing prescription plus the adopted $^{12}$C +$\alpha$
reaction rate are consistent with observations.  We have already shown
that, at least for M5, observational data are in agreement with the
combination of the new reaction rate plus semiconvection without
breathing pulses.

However, CCDW have already shown that current
theoretical predictions for the R parameter, as obtained by using 
the same input physics adopted in the present work,
provide an unrealistically large value for the original He abundance in
galactic GCs. This unpleasant situation could be clarified
only by reducing the uncertainties related to the adopted physical
inputs.

To summarize, the cross section for $^{12}$C +$\alpha$ reaction, the
amount of central mixing in He burning stars, the evaluation of the
original He in globular cluster stars and the ratio N$_{\mathrm AGB}$/N$_{\mathrm HB}$
are strongly connected.  In this context, better evaluations of one or
more out of the quoted quantities would be of great relevance to
assess the problem on a more firm basis.

\section{Acknowledgments}

It is a pleasure to acknowledge G. Bono for a careful reading
of the manuscript. We thank the anonymous referee for useful comments.

\label{lastpage}


\begin{thebibliography}{9}

%\bibitem{} Alves D.R., Sarajedini A. 1999, ApJ 511, 225
\bibitem{} Angulo C., Arnould M., Rayet M. et al. (NACRE collaboration), 1999, Nuclear Physics A 656,3
%\bibitem{} Bressan A., Bertelli G. \& Chiosi C., 1986, Mem. Sait 57, 411
\bibitem{} Buchmann L., 1996, ApJ 468, L127
\bibitem{} Buonanno R., Corsi C.E., Fusi Pecci F. 1985, A\&A 145, 97
\bibitem{} Caputo F., Castellani V., Chieffi A., Pulone L. \& Tornambe A., 1989, ApJ 340, 241
\bibitem{} Carney B.W., 1996, PASP 108, 900
\bibitem{} Carretta E., Gratton R.G., 1997, A\&AS 121, 95
%\bibitem{} Cassisi S., Salaris M., 1997, MNRAS 285, 593
\bibitem{} Cassisi S., Castellani V., Degl'Innocenti S., Weiss A., 1998,
A\&AS 129, 267 (CCDW)
\bibitem{} Cassisi S., Castellani V., Degl'Innocenti S., Salaris, M., Weiss A., 1999,
A\&AS 134, 103
\bibitem{} Castellani V., Degl'Innocenti S., 1999, A\&A 344, 97
\bibitem{} Castellani V., Chieffi A., Pulone L., Tornambe A., 1985, ApJ 296, 204
\bibitem{} Castellani V., Chieffi S., Pulone L., 1991, ApJS 76, 911 (CCP)
\bibitem{} Castellani V., Degl'Innocenti S., Marconi M., 1999, A\&A 349, 834
%\bibitem{} Castellani V., Degl'Innocenti S., Girardi L. et al. 2000, A\&A 354, 150
\bibitem{} Castelli F., Gratton R.G., Kurucz, R., 1997, A\&A 318, 841
\bibitem{} Caughlan G.R., Fowler W.A., 1988, Atom. Data Nucl. Data Tables 40, 283
\bibitem{} Caughlan G.R., Fowler W.A., Harris M.J., Zimmerman B.A., 1985, Atom. Data Nucl. Data Tables 32, 197 
\bibitem{} Chaboyer B., 1995, ApJ 444, L9
\bibitem{} Chieffi A., Straniero O., 1989, ApJS 71, 47
\bibitem{} D'Antona F., Mazzitelli I., 1990, ARAA 28, 139
\bibitem{} Dorman B., Rood R.T., 1993, ApJ 409, 387
\bibitem{} Gratton R.G., Fusi Pecci F., Carretta E. et al. , 1997, ApJ 491, 749
\bibitem{} Haft M., Raffelt G., Weiss A., 1994, ApJ 425, 222
%\bibitem{} Harris, W.E., 1996, AJ, 112, 1487
\bibitem{} Hubbard W.B., Lampe M.,  1969, ApJS 18, 297
\bibitem{} Huebner W.F., Merts, A.L., Magee N. H., Argo M. F., 1977,
Los Alamos Sci. Lab. Rept. (LA-6760-M)
%\bibitem{} Iben I.Jr., 1968, Nature 220, 143
\bibitem{} Iglesias C.A., Rogers F.J., 1996, ApJ 464, 943
\bibitem{} Itoh N., Mitake S., Iyetomi H., Ichimaru S., 1983, ApJ 273, 774
%\bibitem{} Kraft R.P., Sneden C., Langer G.E., Shetrone M.D., Bolte M., 1995, AJ 109, 2586
\bibitem{} Munakata H. Kohyama Y., Itoh N., 1985, ApJ 296, 197
\bibitem{} Pagel B.E.J., Portinari L., 1998, MNRAS 298, 747
\bibitem{} Proffitt C.R., VandenBerg D.A., 1991, ApJS 77, 473
\bibitem{} Pulone L., 1992, Mem. SAIt 63, 485
\bibitem{} Rogers F.J., 1994 in "The equation of state in astrophysics",
IUA Colloq. 147, Chabrier G., Schatzman E.L. eds., Cambridge University
Press, Cambridge, p. 16
\bibitem{} Rogers F.J., Swenson F.J., Iglesias C.A., 1996, ApJ 456, 902
\bibitem{} Salaris M., Cassisi S., 1996, A\&A 305, 858
\bibitem{} Salaris M., Weiss A., 1998, A\&A 335, 943
\bibitem{} Salaris M., Chieffi A., Straniero O., 1993, ApJ 414, 580
\bibitem{} Salaris M., Dominguez I., Garc\'ia-Berro E. et al. 1997, ApJ 486, 413
%\bibitem{} Sandage A., 1990, ApJ 350, 603
\bibitem{} Sandquist E.L., Bolte M., Stetson P.B., Hesser J.E., 1996, ApJ 470, 910
\bibitem{} Schlegel D.J., Finkbeiner D.P., Davis M., 1998, ApJ 500, 525
\bibitem{} Sneden C., Kraft R.P., Prosser C.F., Langer G.E.,1992, AJ 104, 2121
\bibitem{} Sosin C., Piotto G., Djorgovski S.G. et al., 1996, Proceeding of the workshop "Stellar Ecology",
Marciana Marina, Italy, Rood R.D. and Renzini A. eds., p.92
\bibitem{} Straniero O., 1988, A\&AS 76, 157
\bibitem{} Sweigart A., 1990,  in ``Confrontation between stellar pulsation and evolution''; Proceedings of the Conference,
           Bologna, Italy, May 28-31. San Francisco, CA, Astronomical Society
           of the Pacific, 1990, p. 1.
\bibitem{} Sweigart A.V., Demarque P., 1972, A\&A 20, 445
\bibitem{} Sweigart A.V., Demarque P., 1973, in "Variable Stars in Globular Clusters"'
ed. J.D. Fernie (Dordrecht:Reidel), p. 221 
%\bibitem{} Sweigart A., Renzini A., Tornamb\'e A., 1987, ApJ 312, 762
%\bibitem{} VandenBerg D.A., 2000 ApJ, {\sl in press}
\bibitem{} Wood M.A., 1992, ApJ 386, 539
\bibitem{} Zoccali M., Cassisi S., Bono G. et al.  2000, ApJ 538, 289
\end{thebibliography}
\end{document}